\documentclass[preprint]{jpsj2}
\usepackage{psfig}

\newlength{\minitwocolumn} 
\def\runtitle{Stress and Strain in Flat Piling of Disks}
\def\runauthor{Shio Inagaki and James T. Jenkins}

\title{ Stress and Strain in Flat Piling of Disks }
\author{ Shio Inagaki\thanks{%
E-mail address: shio@mail.ne.jp,
Present Address:
Department of Basic Science,
University of Tokyo, Komaba, Tokyo 153-8902, Japan
} and James T. Jenkins}

\inst{%
Department of Theoretical and Applied Mechanics\\
Cornell University, Ithaca, NY 14853 USA{\large \ }}

\abst{%
We have created a flat piling of disks in a numerical experiment using the
Distinct Element Method (DEM) by depositing them under gravity. In the
resulting pile, we then measured increments in stress and strain that were
associated with a small decrease in gravity. We first describe the stress in
terms of the strain using isotropic elasticity theory. Then, from a
micro-mechanical view point, we calculate the relation between the stress
and strain using the mean strain assumption. We compare the predicted values
of Young's modulus and Poisson's ratio with those that were measured in the
numerical experiment.
}

\kword{%
granular piling, stress-strain relation, Distinct Element Method
}

\begin{document}
\sloppy
\maketitle


\section{Introduction}

\label{sec:intro}

The ultimate goal of our research is to describe how stresses propagate
through granular media. Physical experiments \cite{Sarero} indicate that
when a localized force is applied at a surface of a layer of sand, the
stress distribution on the bottom of the layer depends on how the layer was
constructed. The pressure distribution under a conical sand pile also
depends on the way the pile was built and may have a local minimum under the
apex of a pile \cite{Vanel,Geng}. We don't yet know how to describe the way
that stress propagates through a pile or how the propagation depends on its
construction history.

In this paper, as a first step, we focus on a two-dimensional example with
possible anisotropy in the vertical direction: a flat pile of circular,
elastic, frictional disks, deposited under gravity onto a frictional, flat
base. In this case, as shown in \S~\ref{subsec:contact}, we find that the
anisotropy due to the gravity is small enough to neglect. In \S~\ref%
{sec:elast}, we write down equations of force balance, assuming that the
pile is continuous body, in order to predict stress and strain within it.
Assuming that the pile is an isotropic elastic body, these equations are
solved exactly. But the problem is that material constants, such as the
Young's modulus and Poisson's ratio are unknown. In the next chapter, we adopt
the mean strain assumption in order to predict the material constants of an
assembly of disks in terms of micro-scopic material properties, such as the
contact stiffness. Here, a relationship between the micro-scopic and
macro-scopic material properties is obtained. In \S~\ref{sec:dem} we
build a flat pile with elastic, frictional circular disks in a numerical
experiment using the Distinct Element Method, a kind of molecular dynamics.
Then, the stress and strain in the pile are measured in a deformation
produced by reducing gravity, and the predictions that are made in \S~\ref%
{sec:meanstrain} are tested against the measured values. In the last
chapter, 
we discuss the reasons for the differences between the predictions based on
the mean strain assumption and the measured results and indicate what more
should be done.


\section{Elasticity theory}

\label{sec:elast}


\subsection{Continuum theory}

\label{subsec:elast}

The equation of motion for a granular piling is written as 
\[
\frac{\partial \sigma _{ij}}{\partial x_{j}}+\rho \ b_{i}=\rho \ a_{i},
\]%
where $\sigma _{ij}$ is the stress, $\rho $ is the mass density, $b_{i}$ is
the body force per unit area, and $a_{i}$ is the acceleration. In this
paper, we focus on only a two-dimensional case, so the summation should be
taken from $1$ to $2$. The body force is gravity having only a vertical
component, $b_{2}=-g$.

We are interested in a static piling, so $a_{i}=0$, $i=1,2$. The equilibrium
equations for the two-dimensional case are 
\begin{eqnarray}
\frac{\partial \sigma _{11}}{\partial x_{1}}+\frac{\partial \sigma _{12}}{%
\partial x_{2}} &=&0,  \label{eq:x} \\
\frac{\partial \sigma _{12}}{\partial x_{1}}+\frac{\partial \sigma _{22}}{%
\partial x_{2}}-\rho \ g &=&0.  \label{eq:y}
\end{eqnarray}

The boundary conditions are defined 
for a pile of uniform height $h$: the horizontal boundary is periodic and,
at the free surface, the shear stresses and the vertical stress are zero, $%
\sigma _{22}(x_{2}=h)=0$ and $\sigma _{12}(x_{2}=h)=0.$ Then, by symmetry $%
\sigma _{12}\equiv 0$. By solving eqs.~(\ref{eq:x}) and (\ref{eq:y}), the
components of stresses are 
\begin{eqnarray}
\sigma _{11} &=&\sigma _{11}(x_{2}),  \label{eq:strx} \\
\sigma _{22} &=&\rho \ g(x_{2}-h).  \label{eq:stry}
\end{eqnarray}%
In these equations, $\sigma _{11}$ is indeterminate, so the exact solutions
can not be obtained using only this information. In the next section, we
will adopt isotropic elasticity theory in order to obtain an exact
determination of $\sigma _{11}$.


\subsection{Isotropy}

\label{subsec:iso}

The horizontal and vertical strains, $E_{11}$ and $E_{22},$ can be obtained
by introducing the constitutive relations of isotropic elasticity: 
\begin{eqnarray}
E_{11} &=&\frac{1}{E}(\sigma _{11}-\nu \sigma _{22}),  \label{eq:hookex} \\
E_{22} &=&\frac{1}{E}(\sigma _{22}-\nu \sigma _{11}),  \label{eq:hookey}
\end{eqnarray}%
where $\nu $ is Poisson's ratio and $E$ is Young's modulus.

The horizontal stress, $\sigma _{11}$, given by eq.~(\ref{eq:strx}), is
obtained using eq.~(\ref{eq:hookex}) and requiring that $E_{11}\equiv 0$
because of the horizontal periodic boundary condition. Then 
\begin{eqnarray}
\sigma _{11} &=&\nu \sigma _{22}  \nonumber \\
&=&\nu \rho \ g(x_{2}-h).  \label{eq:hooksx2}
\end{eqnarray}%
Substituting eqs~(\ref{eq:stry}) and (\ref{eq:hooksx2}) into eq.~(\ref%
{eq:hookey}), the vertical strain is 
\begin{eqnarray}
E_{22} &=&\frac{1-\nu ^{2}}{E}\sigma _{22}  \nonumber  \\
&=&\frac{1-\nu ^{2}}{E}\rho \ g(x_{2}-h).  \label{eq:ey}
\end{eqnarray}


\section{Prediction of the material constants}

\label{sec:meanstrain}

From a micro-mechanical view point, the average stress
can be written using an orientational distribution of contacts, $D(%
\mathbf{{n})}$, and a contact force, $\mathbf{f}^{c}$, as 
\[
\sigma _{ij}=\frac{\gamma }{\pi a}\int D(\mathbf{n}){f}_{i}^{c}n_{j}d\Omega
, 
\]%
where $\gamma $ is the area fraction of the disks, $a$ is a radius of disk, $%
\mathbf{n}$ is the unit vector from the center of the disk to a contact, and 
$d\Omega $ is the element of contact angle.

Here, we make the strong assumption that the contact displacement is
determined by the average strain as 
\[
u_{p}=aE_{pq}n_{q}.
\]%
Then, the contact force is also determined by the average strain as 
\begin{eqnarray*}
{f}_{i}^{c} &=&K_{ip}u_{p} \\
&=&K_{ip}aE_{pq}n_{q},
\end{eqnarray*}%
where 
\[
K_{ip}=K_{N}n_{i}n_{p}+K_{T}(\delta _{ip}-n_{i}n_{p}),
\]%
in which $K_{N}$ and $K_{T}$ are normal and tangential stiffness,
respectively. 

We assume that the orientational distribution of contact is isotropic; then 
\begin{eqnarray}
D(\mathbf{{n})=}\frac{k}{2\pi }\mathbf{,} \label{eq:contact}
\end{eqnarray}%
where $k$ is the average number of contacts per disk. \cite{Jenkins}. 
As shown in \S~\ref{subsec:contact}, it is appropriate to assume that the
contact angle distribution is isotropic.

Then, 
\begin{eqnarray}
\sigma _{ij}=\frac{\gamma }{\pi a}\int \frac{k}{2\pi }aK_{ip}n_{q}n_{j}d%
\Omega E_{pq}. \label{eq:sigma}
\end{eqnarray}
The general form of Hooke's law is 
\begin{eqnarray}
\sigma _{ij}=C_{ijpq}E_{pq}. \label{eq:gen_hooke}
\end{eqnarray}
By comparing eqs~(\ref{eq:sigma}) and (\ref{eq:gen_hooke}), 
\[
C_{ijpq}=\frac{\gamma }{\pi a}\int \frac{k}{2\pi }aK_{ip}n_{q}n_{j}d\Omega . 
\]
After the integration in eq.~(\ref{eq:gen_hooke}) is carried out, eq.~(\ref%
{eq:sigma}) becomes 
\[
\sigma _{ij}=\frac{\gamma k}{4\pi }[(K_{N}+K_{T})E_{ij}+\frac{1}{2}%
(K_{N}-K_{T})E_{kk}\delta _{ij}]. 
\]

When the stress is expressed in terms of strain, the coefficients are called
Lam\'{e}'s constants. Now, Lam\'{e}'s constants are obtained in terms of
stiffness of disks as 
\begin{eqnarray*}
2\mu &=&\frac{\gamma k}{4\pi }(K_{N}+K_{T}), \\
\lambda &=&\frac{\gamma k}{8\pi }(K_{N}-K_{T}).
\end{eqnarray*}%
Then Young's modulus, $E$ and Poisson's ratio, $\nu $, are converted from Lam%
\'{e}'s constants, so they can also be expressed in terms of stiffness of
disks as 
\begin{eqnarray}
\nu &=&\frac{\lambda }{2\mu +\lambda }=\frac{K_{N}-K_{T}}{3K_{N}+K_{T}},%
\nonumber \\
E &=&\frac{4\mu (\mu +\lambda )}{2\mu +\lambda }=\frac{\gamma k}{\pi }\frac{%
(K_{N}+K_{T})K_{N}}{3K_{N}+K_{T}}.  \label{eq:E}
\end{eqnarray}%
Using this relation between micro-scopic and macro-scopic material property,
we can predict Young's modulus and Poisson's ratio.


\section{Numerical experiments}

\label{sec:dem}

In this paper, we adopt Distinct Element Method (DEM), which was invented by
Cundall\cite{Cundall}, in order to produce deformed granular aggregates in
which stress and strain can be measured. We follow the algorithm of A.
Shimosaka \cite{granular} and H. Hayakawa \cite{Hayakawa}.

Using DEM, we make a two-dimensional flat pile that is composed of elastic,
frictional, circular disks on a frictional smooth bottom, and measure stress
and strain at each point.

\subsection{Setting}

\begin{figure}[h]
\caption{}
\label{fig:pos}\centerline{
\includegraphics[width=150mm]{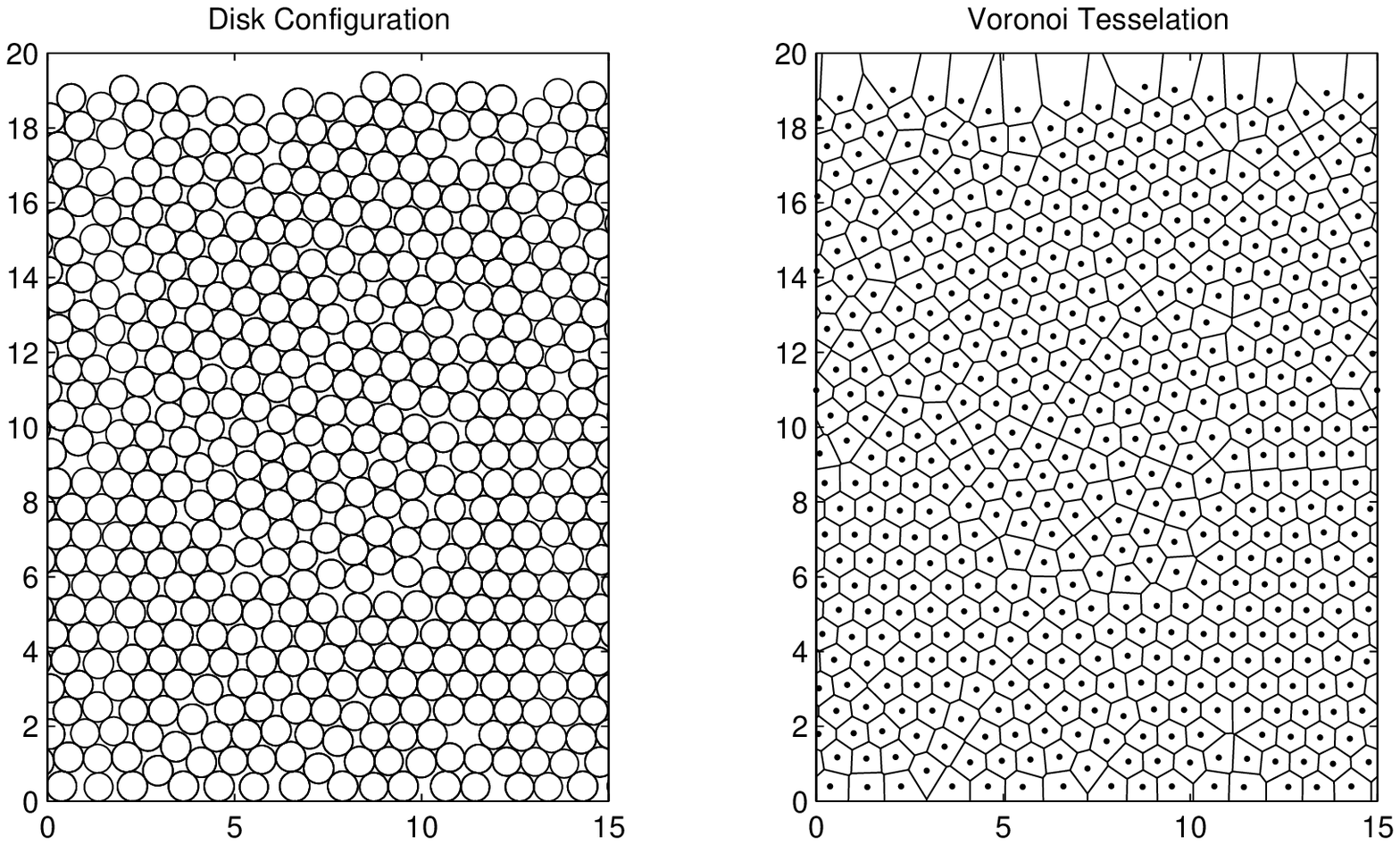}
}
\end{figure}

Elastic, frictional, circular disks are deposited on a flat frictional
bottom, layer by layer. Two diameters of disks are chosen that are slightly
different from each other, in order to avoid a crystal structure. The
horizontal boundary is periodic. We let this pile relax until the disks come
to a static state, after dissipating all of their kinetic energy in
collisions. (See the left panel of Fig.~\ref{fig:pos}.) There is the
possibility of the existence of depositional anisotropy in the vertical
direction that will be considered later in \S~\ref{subsec:contact}. The
right panel of Fig.~\ref{fig:pos} shows a Voronoi tessellation of this pile.
Voronoi tessellation divides the whole domain into cells using perpendicular
bisectors of the lines of centers, so that 
each cell contains one center, indicated by a dot. Voronoi cells are used to
measure strain in a granular assemblies as defined in \S~\ref{subsec:str}.

The parameters that are used in the actual simulation are normalized so that
the maximum disk diameter, the gravitational acceleration, and the mass per
unit area are all unity. Consequently, upon taking the summation over all
contacts $B$ with disk $A$, the dimensionless equation of motion for disk $A$
is 
\[
m_{A}^{\prime }\frac{d^{2}x_{A}^{\prime }}{dt^{\prime }{}^{2}}%
=\sum\limits_{B}\left[ \eta ^{\prime }\frac{d(x_{B}^{\prime }-x_{A}^{\prime
})}{dt^{\prime }}-k^{\prime }(x_{B}^{\prime }-x_{A}^{\prime })\right]
-m_{A}^{\prime },
\]
where $x=lx^{\prime }$, $t=\sqrt{l/g}\ t^{\prime }$. In our model, contact
forces consist of elastic and viscous forces which are linearly proportional
to the relative displacement and the relative velocity, respectively. A dash
denotes a non-dimensional variable. For example, a dimensionless elastic
coefficient and a dimensionless viscous coefficient are calculated as $%
k^{\prime }=(l/\bar{m}g)k$ and $\eta ^{\prime }=(1/\bar{m})\sqrt{l/g}\ \eta ,
$ respectively, where $\bar{m}$ is the mass per unit area and $m_{A}^{\prime
}$ is the dimensionless mass of disk $A$. The dimensional parameters used in
our calculations are shown in Table~\ref{table:param}.

\begin{table}[t]
\caption{
Parameters in our numerical experiments
}
\label{table:param}
\begin{center}
\begin{tabular}
{@{\hspace{\tabcolsep}\extracolsep{\fill}}cc} 
\hline
Diameters of disks  & $7.6$ and $8.0$ ($10^{-3}m$) \\ \hline
Thickness of disk & $6.0$ $(10^{-3}m)$ \\ \hline
Density of disk & $1.06\times 10^{3}$ $(kg/m^{3})$ \\ \hline
Gravitational Accel. & $10$ $(m/s^{2})$ \\ \hline
Restitution Coef. & $0.6$ \\ \hline
Frictional Coef. & $0.4$ \\ \hline
Normal elastic Coef.($k_{n}$) & $1.27\times 10^{4}$ $(N/m)$ \\ \hline
Tang. elastic Coef.($k_{t}$) & $2.54\times 10^{3}$ $(N/m)$ \\ \hline
Normal viscous Coef.($\eta _{n}$) & $1.00$ $(kg/s)$ \\ \hline
Tang. viscous Coef.($\eta _{t}$) & $1.00$ $(kg/s)$ \\ \hline
Time step ($dt$) & $3.16\times 10^{-7}$ $(s)$ \\ \hline
\end{tabular}
\end{center}
\medskip
\end{table}

\subsection{Measurement of stress and strain}

\label{subsec:str}

There exist several definitions for stress and strain of granular
aggregates. \cite{Rothenburg,Bagi,Satake} In this paper we adopt a
simplified form of definitions \cite{Rothenburg,Bagi}, that are consistent
with each other in two dimensions. After reducing gravity by $10\%$, we
measure the displacements of the disks and the increments of the contact
forces and, from them, calculate increments of stress and strain. The
increment of 
the strain at a point at the center of disk $A$ is taken to be%
\[
\dot{E_{ij}}^{A}=\frac{\gamma }{\pi a_{A}^2}\sum\limits_{B}\dot{u_{i}}%
^{AB}n_{j}^{AB}l^{AB},
\]%
where $\gamma $ is the area fraction of the diska,  $\mathbf{{u}^{AB}}$ is a
displacement of disk $A$ relative to disk $B$, $\mathbf{{n}^{AB}}$ is the
unit vector in the direction of contact between disk $A$ and $B$, and $l^{AB}
$ is the length of the side of a Voronoi cell which is shared by cell $A$
and cell $B$, with the name of a cell the same as the name of a disk which
is inside of the cell. The corresponding increment in stress is taken to be%
\[
\dot{\sigma _{ij}}^{A}=\frac{\gamma }{\pi a_{A}}\sum\limits_{C}\dot{{f_{c}}%
_{i}}^{AC}n_{j}^{AC},
\]%
where $\mathbf{{f_{c}}^{AC}}$ is a contact force exerted on disk $A$ by disk 
$C$ and $a_{A}$ is the radius of disk $A$. In the definition of strain, the
summation is taken over all the neighbourings which share sides of Voronoi
cells in between; while for stress, the sum is taken over all of the pairs of
contacts.

\begin{figure}[h]
\begin{center}
\caption{}
\includegraphics[width=4in]{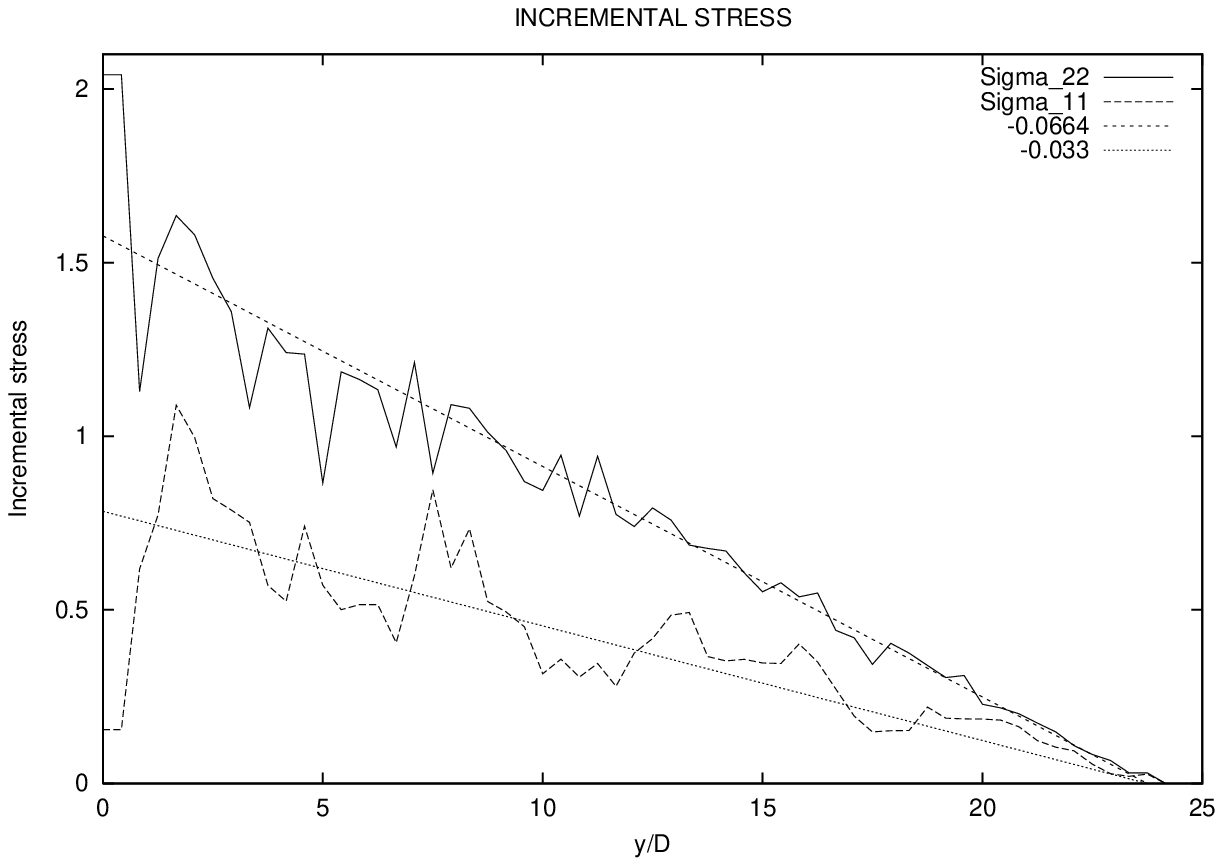}
\label{fig:stress}
\end{center} 
\end{figure}

The increments of stress and strain are measured at each disk center
according to the definitions above. They are taken to be an average over
disks which are included in a horizontal slice with a width of $1.5$
dimensionless units at each height. Figure~\ref{fig:stress} shows the
increment of the dimensionless stress versus. the height, normalized by the
maximum diameter. It is linearly proportional to the height of the pile due
to the weight of the disks. %

Now we are interested in the increments of stress and strain associated with
reduction of gravity by $10\%$. Then the increments of stress components are
expressed in terms of the increment of gravity, $\delta g=-0.1g$ as 
\begin{eqnarray}
\dot{\sigma _{11}} &=&\nu \ \dot{\sigma _{22}}  \label{eq:hooksx} \\
&=&\nu \rho \ \delta g(x_{2}-h)  \label{eq:strx2}
\end{eqnarray}%
and 
\begin{equation}
\dot{\sigma _{22}}=\rho \ \delta g(x_{2}-h). \label{eq:stry2}
\end{equation}
The slope of the line along $\dot{\sigma _{22}}$ coincides well with the
value, $-0.07$, which is estimated from eq.(\ref{eq:stry2}), where $\rho
=0.83 $ which is calculated from the area fraction. In this respect, the
continuous description pictures the stress well. 

\begin{figure}[h]
\begin{center}
\caption{}
\includegraphics[width=4in]{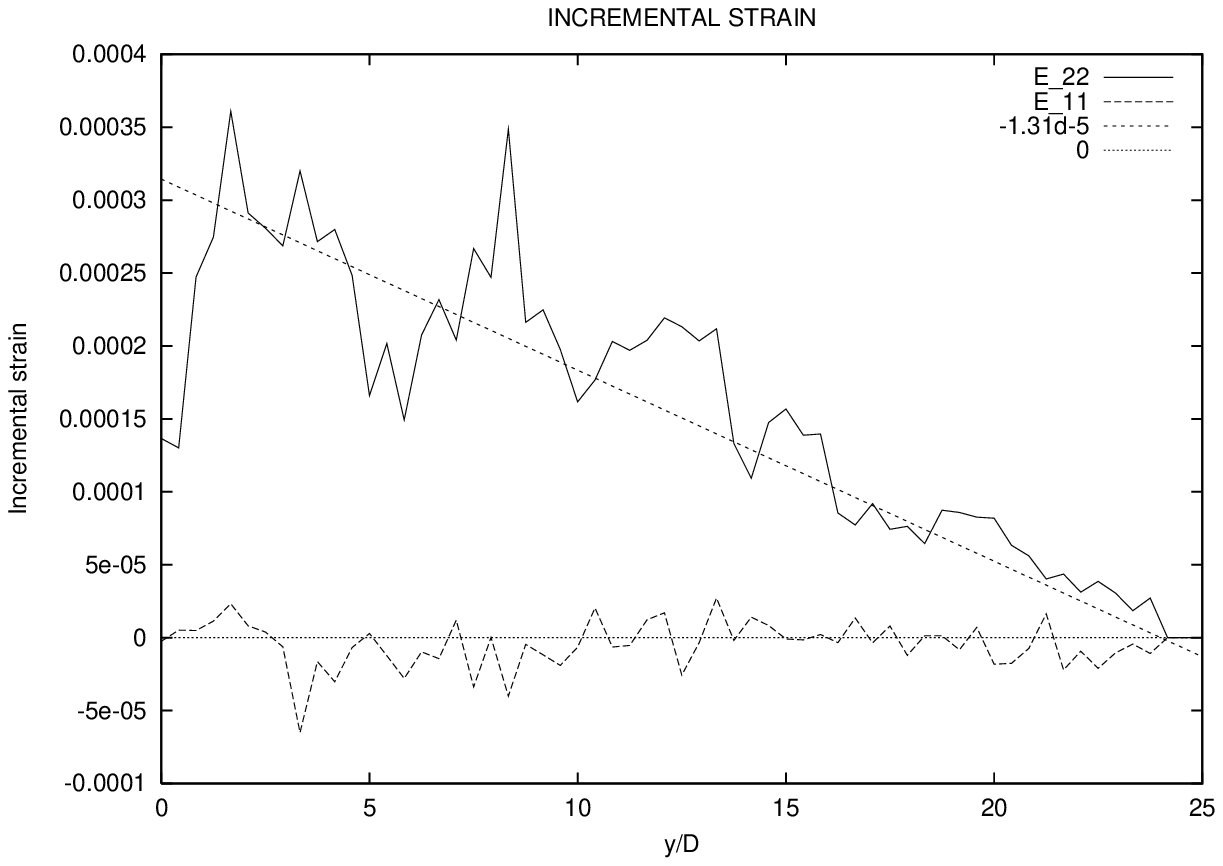}
\label{fig:strain} 
\end{center} 
\end{figure}

Figure~\ref{fig:strain} shows the increment of strain versus. height. From eq.(%
\ref{eq:ey}), the increment of strain is 
\begin{eqnarray}
\dot{E_{22}} &=&\frac{1-\nu ^{2}}{E}\dot{\sigma _{22}}  \label{eq:ey3} \\
&=&\frac{1-\nu ^{2}}{E}\rho \ \delta g(x_{2}-h).  \label{eq:ey2}
\end{eqnarray}%
Comparing eq.~(\ref{eq:ey2}) and Fig.~\ref{fig:strain}, we found that $%
E_{22} $ is related linearly with the height of the pile with some
fluctuation.

Next, let us consider the relation between stress and strain. Figure~\ref%
{fig:se} shows the increment of the dimensionless stress versus the
increment of strain. From eq.~(\ref{eq:ey3}), stress is expected to be
linearly proportional to strain. The linear relation can be seen in Fig.~\ref%
{fig:se} between stress and strain in our numerical experiment.

\subsection{Contact angle distribution}

\label{subsec:contact} 
Figure~\ref{fig:contact} shows the contact angle
distribution. There are peaks around $\pi /3$, $2\pi /3,$ and $\pi $. It
indicates that the disk configuration is nearly a hexagonal packing.
Although we expected that there would exist a depositional anisotropy due to
the process of construction of the pile, it was found that 
the contact angle distribution can be assumed to be isotropic in a
macroscopic sense. So the assumption that was made in eq.(\ref{eq:contact}) is
supported. The hexagonal structure also can be seen in the Voronoi
tessellation (see the right panel of the Fig.~\ref{fig:pos}, although it
should be noted that a side of Voronoi cells is defined as a bisection of
the nodes connected the centers of a pair of disks, but it doesn't mean that
they are necessarily in contact as shown by the fact that the average number
of contacts is about $4.7$.


\subsection{Measurement of the material constants}

\label{subsec:measure}

Under the assumption that the material is isotropic, we can characterize its
elastic response using only two constants (e.g., the Lam\'{e} constants, or
Young's modulus and Poisson's ratio). However, we don't yet know such
constants for the granular aggregate as a bulk. In other situations
involving anisotropy, two constants are not be sufficient to characterize
the elasticity of a granular material.

For the isotropic material, we can estimate Young's modulus and Poisson's
ratio from eq.~(\ref{eq:hooksx}) and (\ref{eq:ey2}) as 
\begin{eqnarray}
\nu  &=&\frac{\dot{\sigma _{11}}}{\dot{\sigma _{22}}}, \nonumber \\
E &=&(1-\nu ^{2})\frac{\dot{E_{22}}}{\dot{\sigma _{22}}}. \label{eq:young} 
\end{eqnarray}%
The ratio of $\sigma _{11}$ against $\sigma _{22}$ is almost constant as
shown in Fig.~\ref{fig:nu} except at the surface of the pile. 
Also, Young's modulus, estimated from eq.~(\ref{eq:young}), is shown in Fig.~%
\ref{fig:ymod}. The fluctuation in Fig.~\ref{fig:ymod} is found is larger
than that for Poisson's ratio and may not be negligible.

As predicted in \S~\ref{sec:meanstrain}, using $\gamma =0.83$ and $k=4.7$,
as measured in the numerical simulation, the dimensionless Young's modulus $%
E^{\prime }\equiv (l/\bar{m}g)E$ is $9.3\times 10^{3}$ and the Poisson's
ratio is $0.25$. We can compare those values with those obtained in the
numerical experiments shown in Figs.~\ref{fig:nu} and \ref{fig:ymod}. The
predicted Young's modulus is almost $2.58$ times of the measured value;
while the predicted Poisson's ratio is half of the measured value. That is,
the predicted properties of the pile are stiffer than those measured.


\subsection{Fluctuation of strain}

\label{subsec:fluc}

In our micro-mechanical view point, fluctuations in the strain were not
taken into account. Consequently, the predicted material constants were
larger than those measured in the numerical simulation, because the disks in
the simulation 
can translate and rotate in ways different from that predicted by the
average strain in order to 
satisfy force and moment equilibrium. These additional degrees of freedom
allows them to behave more flexibly than predicted. Figure~\ref{fig:fluc}
shows the fluctuation of strain ($\Delta E_{22}/E_{22}$) versus the height.
The fluctuation, $\Delta E_{22}$, is evaluated by the ratio of the absolute
value of the deviation from the mean strain 
to the local value of the mean strain. On average, the strain fluctuation $%
\delta E_{22}$ is $65\%$ of the strain itself, except near the bottom and
the surface of the pile. The fluctuation of strain is very large, especially
near the boundaries. We need to take this fluctuation into account in order
to better describe the relation between stress and strain \cite{Luigi}.


\section{Conclusion}

\label{sec:conc}

In this paper, we have investigated a flat piling of disks in order to
define the relation between stress and strain using a numerical experiment
and a continuum theory. As a continuum description, we adopted isotropic
elasticity theory. In a micro-mechanical approach, we calculated the stress 
using the mean strain assumption and predicted the Young's modulus and the
Poisson's ratio of the pile. The material constants which are measured in
the numerical experiments are smaller than those which were predicted using
mean strain assumption. 
In order to obtain a better description, we need to introduce the
fluctuation of stress and strain in the pile. The disorderness of the disk
configuration may cause this reduction of stiffness as an aggregate as we
can make sure in a pile of disks with a square lattice structure. As a next
step, we can introduce an anisotropy of a pile, and see how stress
propagation depends on it.

An important questions that should be addressed is the extent to which
elasticity theory can be applied to describe unloading of the pile.
Preliminary numerical experiments indicate that particle sliding gives rise
to irreversible behavior for decreases in gravity larger than ten per cent.
The characterization and description of this inelastic behavior will be the
subject of a subsequent paper.


\clearpage
\centerline{Shio Inagaki and James T. Jenkins}

\clearpage
\centerline{Shio Inagaki and James T. Jenkins}

\clearpage
\centerline{Shio Inagaki and James T. Jenkins}

\clearpage
\centerline{Shio Inagaki and James T. Jenkins}
\begin{figure}[h]
\begin{center}
\caption{}
\includegraphics[width=4.0in]{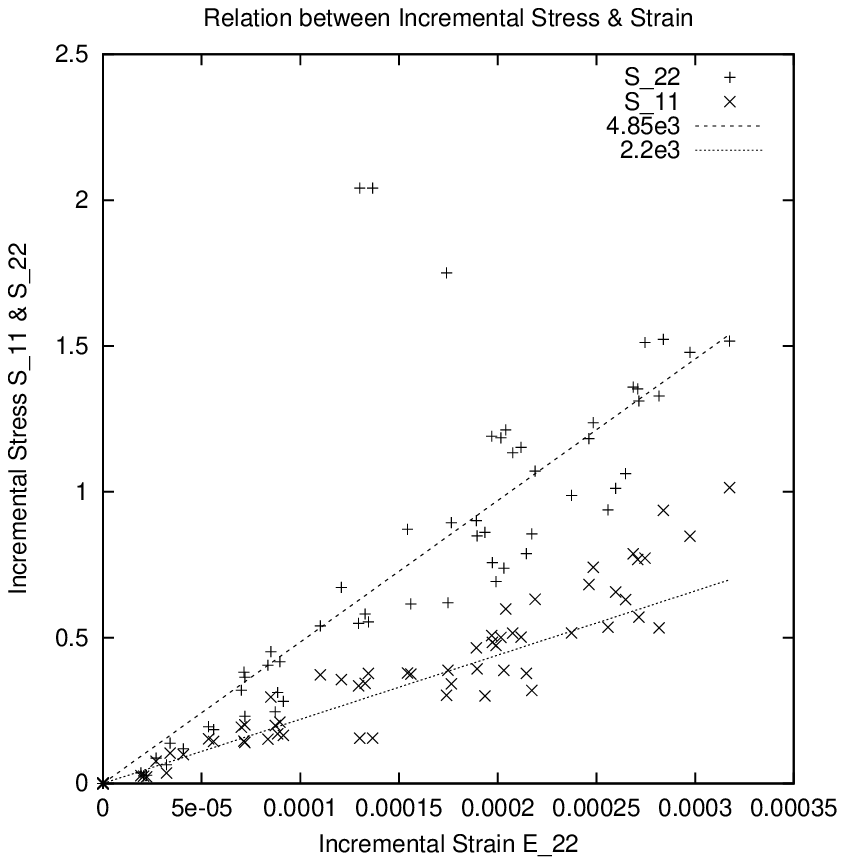}
\label{fig:se} 
\end{center}
\end{figure}
%

\clearpage
\centerline{Shio Inagaki and James T. Jenkins}
\begin{figure}[h]
\begin{center}
\caption{}
\includegraphics[width=4in]{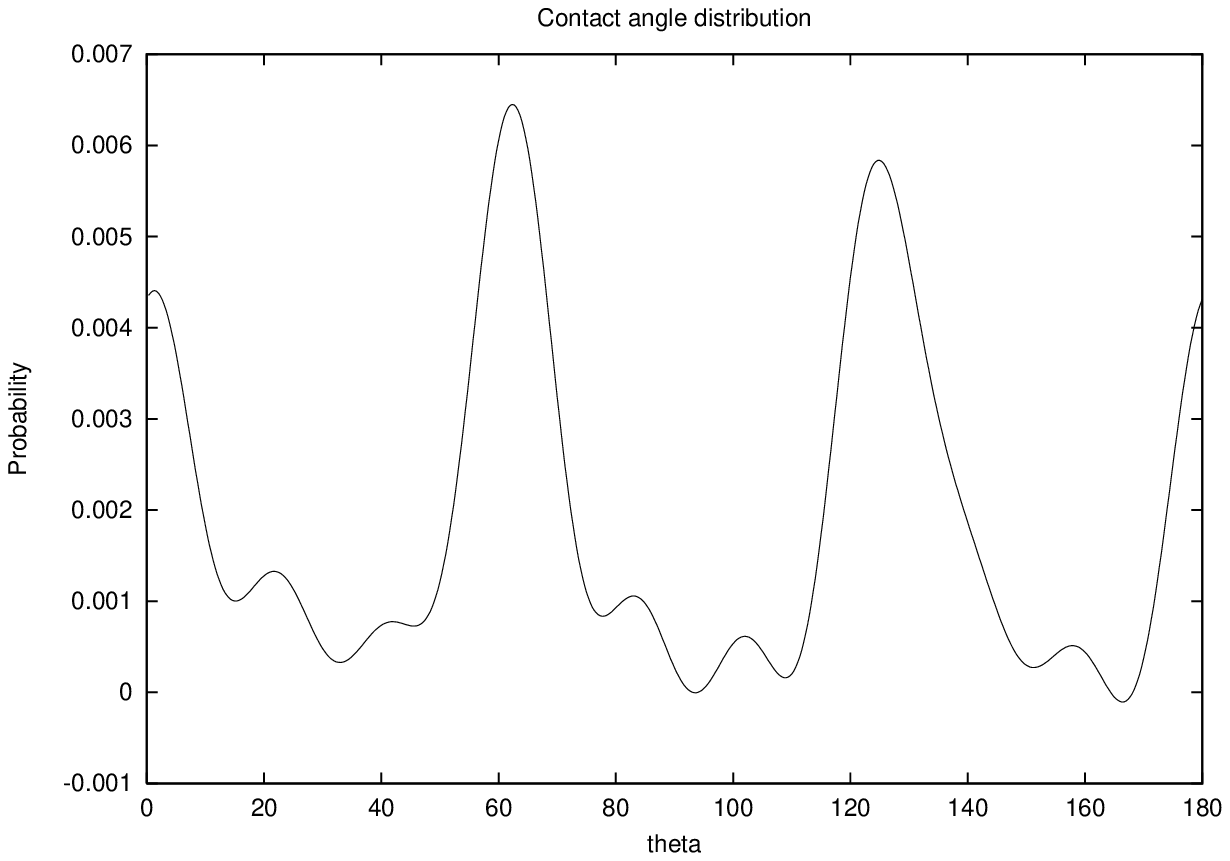}
\label{fig:contact} 
\end{center}
\end{figure}

\clearpage
\centerline{Shio Inagaki and James T. Jenkins}
\begin{figure}[h]
\begin{center}
\caption{}
\includegraphics[width=4in]{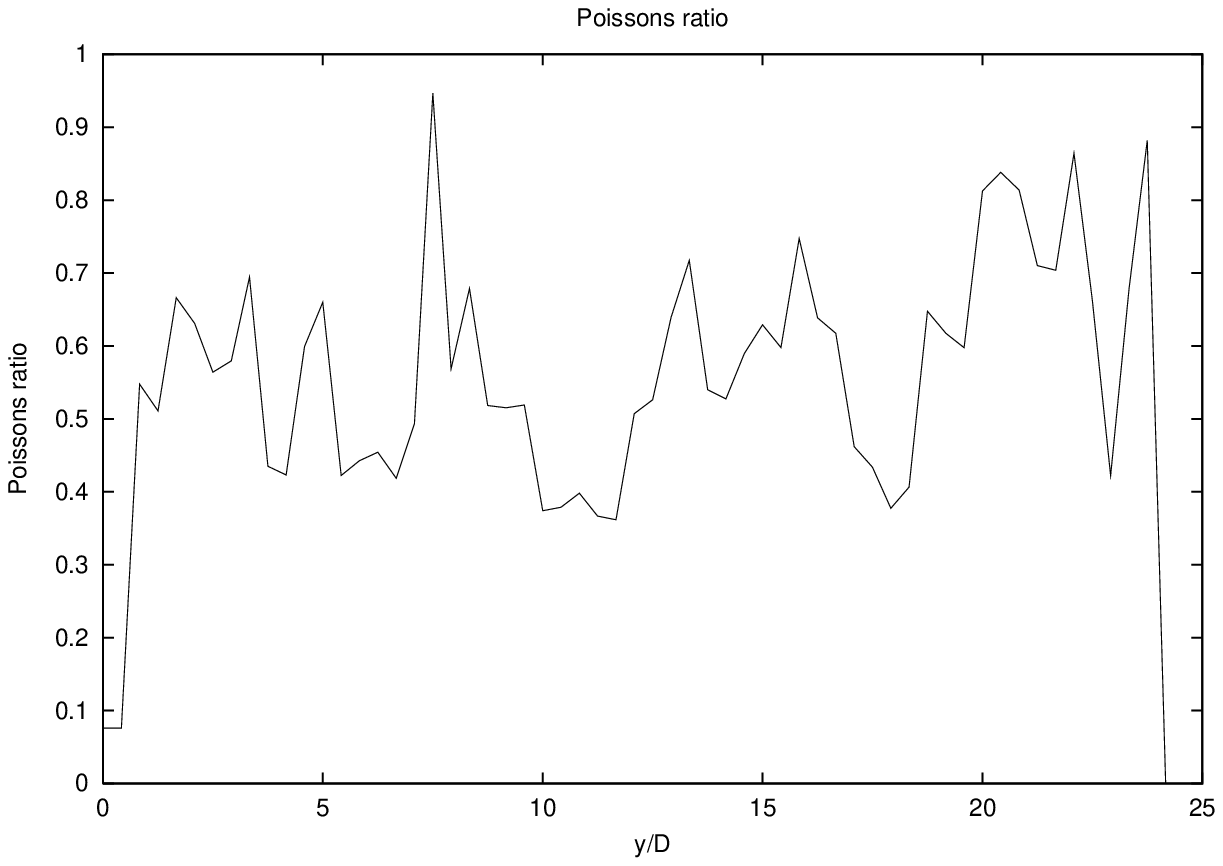}
\label{fig:nu} \end{center}
\end{figure}

\clearpage
\centerline{Shio Inagaki and James T. Jenkins}
\begin{figure}[h]
\begin{center} 
\caption{}
\includegraphics[width=4in]{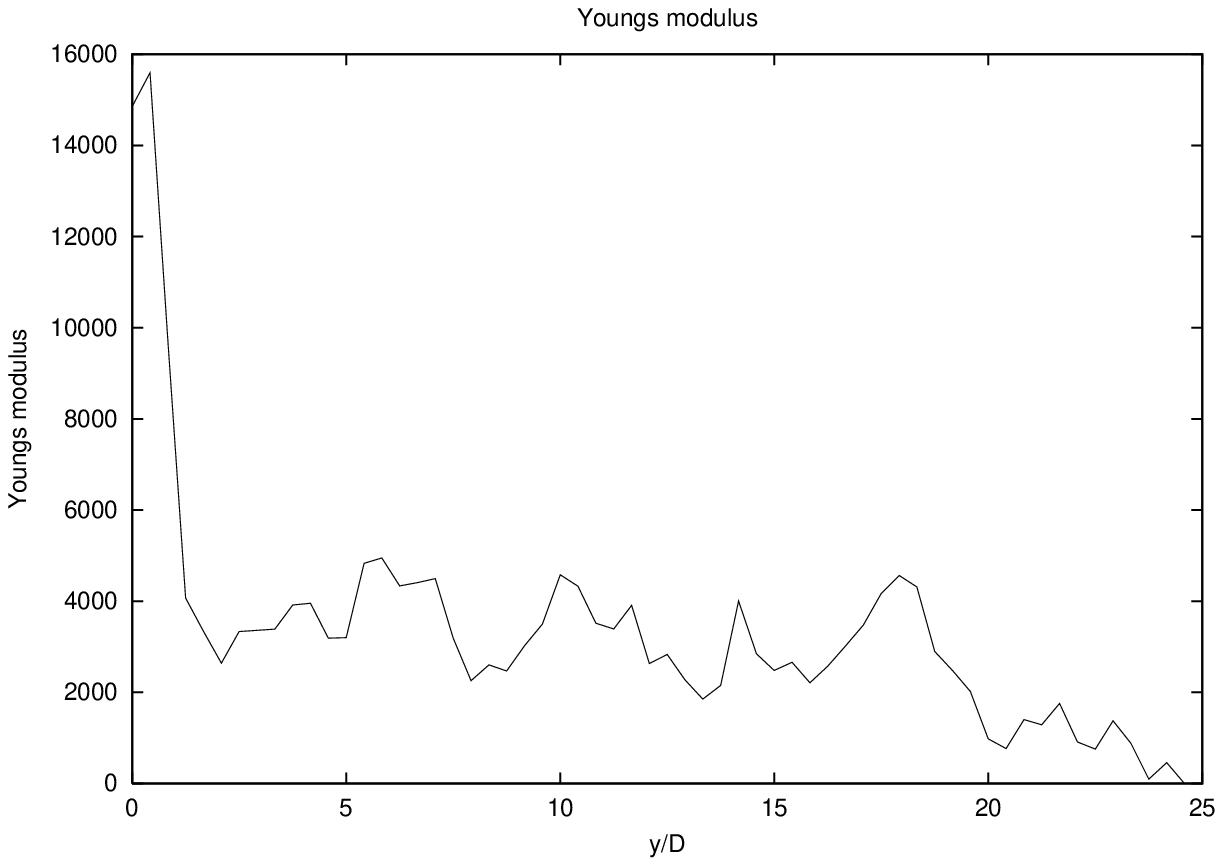}
\end{center} 
\label{fig:ymod}
\end{figure}

\clearpage
\centerline{Shio Inagaki and James T. Jenkins}
\begin{figure}[h]
\caption{}
\centerline{
\includegraphics[width=4in]{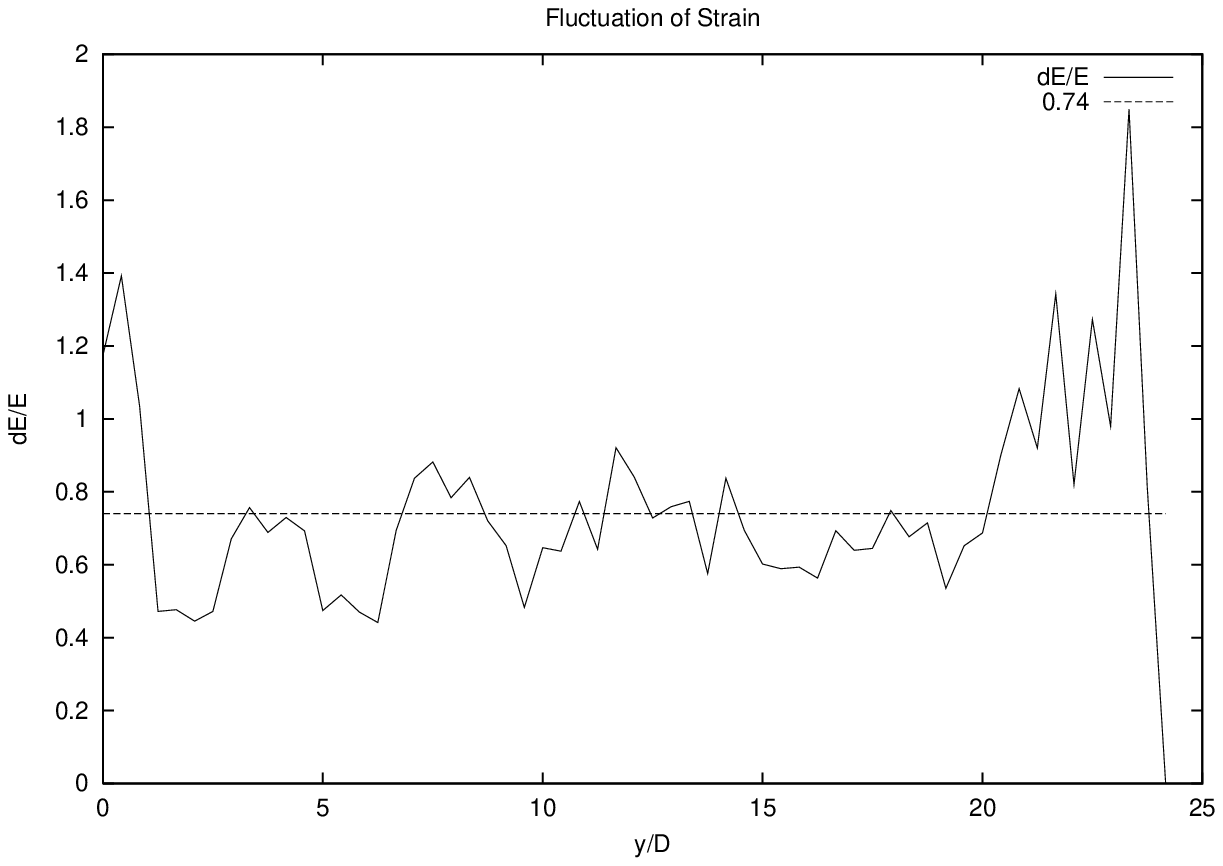}
}
\label{fig:fluc}
\end{figure}

\clearpage
Figure 1.
Left panel: The configuration of disks consisting from 495 disks
in a static state. The horizontal boundary is periodic. The diameters of
disk is $0.76 \ cm$ and $0.8 \ cm$. Right panel: The lines show Voronoi
cells and dots show centers of disks.

\vspace{8mm}
Figure 2.
The increment of the
dimensionless stress versus the height normalized by the maximum diameter (D),
The inclinations of the straight lines for $\sigma_{11} '$ and $\sigma_{22}
'$ are $-0.0664$ and $-0.033$ respectively. This dimensionless stress,
$\sigma '$, is converted to the dimensional value, $\sigma$, as $\sigma =
(\bar{m} g / l) \ \sigma'$.

\vspace{8mm}
Figure 3.
The increment of strain
versus height normalized by the maximum diameter (D). The slopes of the
straight lines for $E_{11}$ and $E_{22}$ are $0$ and $-1.31*10^{-5}$
respectively.

\vspace{8mm}
Figure 4.
The increments of stress,
$\sigma_{xx}$ and $\sigma_{yy}$, versus the increment of strain, $E_{yy}$. The
relation between the increments of stress and strain seems to be linear. The
slope of the straight line is $2.2*10^3$ (for $\sigma_{11} '$ (x),and
$4.85*10^3$ for $\sigma_{22} '$ (+)).

\vspace{8mm}
Figure 5.
Probability distribution
function of contacts versus contact angle from $0$ to $\pi$. Three peaks are
found $\pi/3$,$2 \pi/3$ and $\pi$. It implies that the packing can be
assumed to be almost hexagonal.

\vspace{8mm}
Figure 6.
Poisson's ratio, versus
height normalized by the maximum diameter. This is the ratio of
$\sigma_{11}$ against $\sigma_{22}$ The mean value is about 0.5.

\vspace{8mm}
Figure 7.
The dimensionless Young's modulus, $E'$, versus height normalized by
the maximum diameter.It is estimated from eq.~(\ref{eq:young}). The mean
value is around $3.6*10^3$, and the dimensional value, $E$, is converted
with the relation $E = (\bar{m} g / l) \ E'$.

\vspace{8cm}
Figure 8.
The ratio of the fluctuation of $E_{22}$, $\Delta E_{22}$, to $%
E_{22}$ versus height normalized by the maximum diameter. The strain
fluctuation is almost 74 \% of the strain in average except near the surface
and the bottom.

\end{document}